# Entangling characterization of $(SWAP)^{1/m}$ and Controlled unitary gates

S.Balakrishnan and R.Sankaranarayanan

*Department of Physics, National Institute of Technology,*

*Tiruchirappalli 620015, India.*

We study the entangling power and perfect entangler nature of $SWAP^{1/m}$, for $m \geq 1$, and controlled unitary ($CU$) gates. It is shown that $SWAP^{1/2}$ is the only perfect entangler in the $SWAP^{1/m}$ family. On the other hand, a subset of $CU$ which is locally equivalent to $CNOT$ is identified. It is shown that the subset, which is a perfect entangler, must necessarily possess the maximum entangling power.

## I. INTRODUCTION

Entanglement [1], a fascinating quantum mechanical feature, has been recognized as a valuable resource for quantum information and computation [2]. Much effort has been made to know the production, quantification and manipulation of entangled state [3]. Required information processing can be achieved by the application of appropriate quantum operators (gates) on qubits prepared in a definite state. As two qubit gates have the ability to create entanglement, many research work focus on characterizing the entangling properties of them. The entangling capabilities of a quantum gate are quantified by the entangling power [4] which describes the average entanglement produced by the gate when it is acting on a given distribution of product states. In this description, linear entropy is used to measure the entanglement of a state.

It is well known that the entanglement is a non-local property which is unaffected by local operations. Makhlin introduced local invariants to describe the non-local properties of quantum gates [5]. Two gates are said to be locally equivalent, possessing same local invariants, if they differ only by local operations. Hence, local invariants are convenient measures to identify the local equivalence class of quantum operators. In [6], Zhang et al. showed that the geometric structure of non-local two qubit operations is a 3-Torus. To be precise, every non-local gate is associated with the coordinates of 3-Torus. In terms of the coordinates, it is easy to check whether a gate has the ability to produce maximal entanglement when it acts on some separable states. If the gate produces maximal entanglement then it is known as a perfect entangler [5, 6]. As the maximally entangled states are known to play a central role in the quantum information processing, it is of fundamental importance to identify the perfect entanglers among the non-local gates.

It is known that one $CNOT$ gate can be constructed using two $SWAP^{1/2}$ gates [7]. Since $SWAP^{\alpha}$ family of gates are recognized as the building blocks of universal two qubit gate [8], a detailed understanding of this family is of fundamental importance. In particular we focus on the entangling characterization, complimenting with geometrical representation, of the above family with $\alpha = 1/m$ for $m \geq 1$. Further, we investigate entangling character of another important class of two-qubit gate namely, controlled unitary ($CU$) gates.

Using geometrical representation, it is shown that $SWAP^{1/2}$ is the *only* perfect entangler in the $SWAP^{1/m}$ family. On the other hand, we present a simplified expression for the entangling power and hence obtain conditions for minimum and maximum entangling power of an arbitrary two qubit gate. The simplification led to identify a subset of $CU$ which is locally equivalent to $CNOT$. It is shown that the subset, which is a perfect entangler, must necessarily possess the maximum entangling power as well. In the end, some possible problems emerged from this work are pointed out.

## II. PRELIMINARIES

### (A) Entangling power

The entangling capability of a unitary quantum gate $U$ can be quantified by entangling power (EP) which was introduced by Zanardi et al. [4]. For a unitary operator $U \in U(4)$ the entangling power is defined as

$$EP(U) = \underset{|\psi_1\rangle \otimes |\psi_2\rangle}{average} [E(U|\psi_1\rangle \otimes |\psi_2\rangle)] \qquad (1)$$

where the average is over all product states distributed uniformly in the state space. In the above formula $E$ is the linear entropy of entanglement measure defined as

$$E(|\psi\rangle_{AB}) = 1 - tr(\rho_{A(B)}^2) \qquad (2)$$

where $\rho_{A(B)} = tr_{B(A)}(|\psi\rangle_{AB}\langle\psi|)$ is the reduced density matrix of system $A(B)$. We may note that $0 \leq EP(U) \leq \frac{2}{9}$ [8, 9]. It is to be noted that the linear entropy is related to the well known measure of entanglement namely concurrence [10, 11] through the following expression

$$E(\psi) = \frac{1}{2}C^2(\psi)$$

where the concurrence of a two qubit state $|\psi\rangle_{AB} = \alpha|00\rangle + \beta|01\rangle + \gamma|10\rangle + \delta|11\rangle$ is defined as

$$C(\psi) = 2|\alpha\delta - \beta\gamma|. \qquad (3)$$

While $C = 0$ for product state, it takes the maximum value of 1 for maximally entangled state.

### (B) Perfect Entangler

Two unitary transformations $U, U_1 \in SU(4)$ are called locally equivalent if they differ only by local operations: $U = k_1 U_1 k_2$ where $k_1, k_2 \in SU(2) \otimes SU(2)$ [5]. The local equivalent class of $U$ can be associated with local invariants which are calculated as follows. Any two qubit gates $U \in SU(4)$ can be written in the following form [6, 12, 13]

$$U = k_1 \exp\{\frac{i}{2}(c_1 \sigma_x^1 \sigma_x^2 + c_2 \sigma_y^1 \sigma_y^2 + c_3 \sigma_z^1 \sigma_z^2)\} k_2. \qquad (4)$$

Representing $U$ in the Bell basis:

$$|\Phi^+\rangle = \frac{1}{\sqrt{2}}(|00\rangle + |11\rangle), \quad |\Phi^-\rangle = \frac{i}{\sqrt{2}}(|01\rangle - |10\rangle),$$

$$|\Psi^+\rangle = \frac{1}{\sqrt{2}}(|01\rangle - |10\rangle), \quad |\Psi^-\rangle = \frac{i}{\sqrt{2}}(|00\rangle - |11\rangle)$$

as $U_B = Q^\dagger U Q$ with

$$Q = \frac{1}{\sqrt{2}} \begin{pmatrix} 1 & 0 & 0 & i \\ 0 & i & 1 & 0 \\ 0 & i & -1 & 0 \\ 1 & 0 & 0 & -i \end{pmatrix},$$

the local invariants of the given two qubit gate can be calculated using the formula [6]

$$G_1 = \frac{tr^2(M(U))}{16 \det(U)} \qquad (5a)$$

$$G_2 = \frac{tr^2(M(U)) - tr(M^2(U))}{4 \det(U)} \qquad (5b)$$

where $M(U) = U_B^T U_B$. The relation connecting the local invariants $G_1$, $G_2$ and a point $[c_1, c_2, c_3]$ on 3-Torus geometric structure of non-local two qubit gates are [6]

$$G_1 = \cos^2 c_1 \cos^2 c_2 \cos^2 c_3 - \sin^2 c_1 \sin^2 c_2 \sin^2 c_3 + \frac{i}{4} \sin 2c_1 \sin 2c_2 \sin 2c_3 \qquad (6a)$$

$$G_2 = 4\cos^2 c_1 \cos^2 c_2 \cos^2 c_3 - 4\sin^2 c_1 \sin^2 c_2 \sin^2 c_3 - \cos 2c_1 \cos 2c_2 \cos 2c_3. \qquad (6b)$$

Therefore from the values of $G_1$ and $G_2$ it is possible to find the point on the 3-Torus which corresponding to a local equivalence class of two qubit gates. Employing Weyl group theory to remove the symmetry on the 3-Torus, Zhang et al. [6] have obtained tetrahedron representation (Weyl chamber) of non-local two qubit gates (Fig.1).

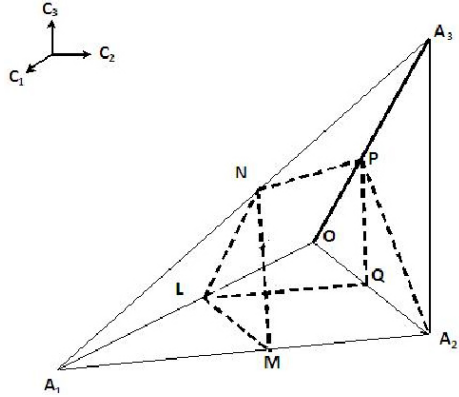

FIG.1: Tetrahedron OA$_1$A$_2$A$_3$, the geometrical representation of non-local two qubit gates, is referred as Weyl chamber. Polyhedron LMNPQA$_2$ (shown in dotted lines) corresponds to the perfect entanglers. Thick line OA$_3$, one edge of the Weyl chamber, corresponds to $SWAP^{1/m}$ gates. The points $L = [\pi/2, 0, 0]$, $A_3 = [\pi/2, \pi/2, \pi/2]$ and $P = [\pi/4, \pi/4, \pi/4]$ correspond to $CNOT$, $SWAP$ and $SWAP^{1/2}$ respectively. The $CNOT$ class of $CU$ lies at the point L and they are all perfect entanglers.

A two qubit gate is called perfect entangler if it can produce a maximally entangled state for some initially separable input state. The theorem for perfect entangler is the following: two qubit gate $U$ is a perfect entangler if and only if the convex hull of the eigenvalues of $M(U)$ contains zero [5, 6]. Alternatively, if the coordinates satisfy the following condition

$$\frac{\pi}{2} \leq c_i + c_k \leq c_i + c_j + \frac{\pi}{2} \leq \pi \quad \text{or} \quad \frac{3\pi}{2} \leq c_i + c_k \leq c_i + c_j + \frac{\pi}{2} \leq 2\pi \qquad (7)$$

where $(i, j, k)$ is a permutation of (1,2,3) then the corresponding two qubit gate is a perfect entangler. Thus the perfect entangler nature of a given two qubit gate $U$ can be ascertained from the corresponding geometric representation.

### III. $SWAP^{1/m}$ FAMILY OF GATES

It is well known that $SWAP$ gate simply interchanges the input states i.e., $SWAP|\psi\rangle|\phi\rangle = |\phi\rangle|\psi\rangle$. Defining

$$SWAP^\alpha = \begin{pmatrix} 1 & 0 & 0 & 0 \\ 0 & \frac{1+\exp(i\pi\alpha)}{2} & \frac{1-\exp(i\pi\alpha)}{2} & 0 \\ 0 & \frac{1-\exp(i\pi\alpha)}{2} & \frac{1+\exp(i\pi\alpha)}{2} & 0 \\ 0 & 0 & 0 & 1 \end{pmatrix}, \qquad (8)$$

it is shown that three such gates with different values of $\alpha$ are the building blocks for the construction of an arbitrary two qubit operations [8]. In such a scheme, $SWAP^\alpha$ can be realized by Heisenberg exchange interaction where $\alpha$ is controlled by adjusting the strength and duration of the interaction. In this section, it is aimed to introduce a family of gates with $\alpha = 1/m$ for $m \geq 1$ and explore their entangling character.

#### (A) Entangling power

We use the following expression to calculate the entangling power of a two qubit gate $U$ [4, 8, 9]:

$$EP(U) = \frac{5}{9} - \frac{1}{36} \{\langle U^{\otimes 2}, T_{1,3} U^{\otimes 2} T_{1,3} \rangle + \langle (SWAP.U)^{\otimes 2}, T_{1,3} (SWAP.U)^{\otimes 2} T_{1,3} \rangle\} \qquad (9)$$

where $\langle A, B \rangle = tr(A^\dagger B)$, referred as Hilbert – Schmidt scalar product and $T_{1,3}$ is the transposition operator defined as $T_{1,3}|a,b,c,d\rangle = |c,b,a,d\rangle$ on four qubit system. The entangling power of $SWAP^\alpha$ is given by [8]

$$EP(SWAP^\alpha) = \frac{1}{12} - \frac{1}{12}\cos(2\pi\alpha) . \qquad (10)$$

For $\alpha = 1/m$,

$$EP(SWAP^{1/m}) = \frac{1}{12}\left(1 - \cos\left(\frac{2\pi}{m}\right)\right) . \qquad (11)$$

It is worth mentioning that $EP(SWAP^{1/2}) = 1/6$, which is the maximum value in the above family. By equating the above entangling power to that of *CNOT*, it is then possible to estimate the number of $SWAP^{1/m}$ gates ($n$) required to simulate *CNOT* for a given value of $m$. Since $EP(CNOT) = 2/9$, the maximum value, we have the following inequality

$$n\{EP(SWAP^{1/m})\} \geq \frac{2}{9} . \qquad (12)$$

Alternatively, the number of gates $n$ for a given $m$ is such that

$$n\left\{1 - \cos\left(\frac{2\pi}{m}\right)\right\} \geq \frac{8}{3} . \qquad (13)$$

Following table shows some integer values of $m$ and the corresponding $n$ that satisfies the above inequality.

| $m$ | 2 | 3 | 4 | 5 | 6 | 7 |
|---|---|---|---|---|---|---|
| No. of gates $n$ | 2 | 2 | 3 | 4 | 6 | 8 |

TABLE 1: Number of $SWAP^{1/m}$ gates ($n$) required for the construction of *CNOT* is shown for few integer values of $m$. The number $n$ is shown to increase with $m$.

Loss et al. [7] have shown that *CNOT* can be constructed using two $SWAP^{1/2}$ gates along with single qubit gates, which is understandable from the point of entangling power. Moreover, since *CNOT* and $SWAP^{1/2}$ possess different local invariants, at least two $SWAP^{1/2}$ gates are needed to simulate $CNOT$. On the other hand, the above table is tempting to conjecture that $CNOT$ can also be constructed using two $SWAP^{1/3}$ gates. Note that the local invariants of $SWAP^{1/3}$ are $G_1 = 0.4063 - 0.1624i$ and $G_2 = 1.5$, which are different from that of $CNOT$. Hence, the later conjecture is well supported by the Makhlin's concept of local equivalence [5]. We conclude this part by pointing that as $m$ increases, the entangling power reaches a maximum of $1/6$ at $m=2$ and decreases for further increase of $m$.

### (B) Perfect Entangler

Expressing $SWAP^{1/m}$ in the Bell basis as $S_B = Q^\dagger SWAP^{1/m} Q$, one can find $M(SWAP^{1/m}) = S_B^T S_B$. Using Eqs. (5a) and (5b) the local invariants are obtained as

$$G_1 = \frac{1}{16}\left[9\exp\left(-\frac{i\pi}{m}\right) + \exp\left(\frac{i3\pi}{m}\right) + 6\exp\left(\frac{i\pi}{m}\right)\right] \tag{14a}$$

$$G_2 = 3\cos\left(\frac{\pi}{m}\right). \tag{14b}$$

Subsequently the geometrical points corresponding to $SWAP^{1/m}$ can be evaluated from Eqs. (6a) and (6b) as

$$[c_1, c_2, c_3] = \left[\frac{\pi}{2m}, \frac{\pi}{2m}, \frac{\pi}{2m}\right], \tag{15}$$

which lie along the line OA$_3$ in the Weyl chamber (see figure.1). For these points, the inequality (7) can be rewritten as

$$1 \leq \frac{2m}{2+m} \leq m \leq 2 \quad \text{or} \quad 2 \geq \frac{2m}{2+m} \geq m \geq \frac{2}{3}.$$

It is easy to convince that the first inequality is satisfied only for $m=2$ and the second inequality is not satisfied for any values of $m$. Hence it is inferred that $SWAP^{1/2}$ is the *only* perfect entangler and the corresponding geometrical point is $P = [\pi/4, \pi/4, \pi/4]$. This is also evident from the fig.1 that the only point P of the line OA$_3$ belongs to the polyhedron of the perfect entanglers. In appendix A, this result is well justified with an explicit calculation of concurrence.

## IV. CONTROLLED UNITARY GATES

In this section we dwell upon the entangling character of another important class two qubit gate namely, the controlled unitary ($CU$) operation:

$$CU = \begin{pmatrix} 1 & 0 & 0 & 0 \\ 0 & 1 & 0 & 0 \\ 0 & 0 & e^{i(\delta+\alpha/2+\beta/2)}\cos(\theta/2) & e^{i(\delta+\alpha/2-\beta/2)}\sin(\theta/2) \\ 0 & 0 & -e^{i(\delta-\alpha/2+\beta/2)}\sin(\theta/2) & e^{i(\delta-\alpha/2-\beta/2)}\cos(\theta/2) \end{pmatrix} \quad (16)$$

where $\alpha, \beta, \theta$ and $\delta$ are real.

### (A) Entangling power

Before calculating the entangling power of $CU$ gates, we make some useful simplification in the expression (9). In what follows, we use the definitions: $A = CU^{\otimes 2}$, $S = SWAP^{\otimes 2}$, $B = (SWAP.CU)^{\otimes 2}$ and $T = T_{1,3}$. Exploiting the property of tensor products [14]: $(A_1 A_2) \otimes (B_1 B_2) = (A_1 \otimes B_1)(A_2 \otimes B_2)$, we can write $B = SA$. With this, we have $\langle B, TBT \rangle = tr(A^+ S^+ TSAT)$ and the entangling power can be rewritten as

$$EP(CU) = \frac{5}{9} - \frac{1}{36}[tr(A^\dagger TAT) + tr(A^\dagger S^\dagger TSAT)]$$

$$= \frac{5}{9} - \frac{1}{36}[tr(A^\dagger TAT + A^\dagger S^\dagger TSAT)]$$

In the last step we use the fact that $tr(A) + tr(B) = tr(A+B)$. It is convenient to rewrite the above expression as

$$EP(CU) = \frac{5}{9} - \frac{1}{36}[tr(A^\dagger RAT)] \quad (17)$$

where $R = T + S^\dagger TS$. From this expression we arrive the conditions for minimum and maximum entangling power of $CU$ gates. The entangling power is minimum i.e., $EP(CU) = 0$, if and only if $tr(A^\dagger RAT) = 20$. Since $tr(RT) = 20$, $EP(CU) = 0$ if $RA = AR$. It is easy to check that the later commutation relation is valid *if* $A$ commutes with $S$ and $T$. That is, if $A$ commutes with $S$ and $T$, the corresponding entangling power is zero. On the other hand, the entangling power is maximum i.e., $EP(CU) = 2/9$, if and only if $tr(A^\dagger RAT) = 12$. In terms of the parameters $\alpha, \beta, \theta$ and $\delta$ the conditions for the minimum and maximum entangling power for $CU$ gate can be expressed as follows. Since

$$tr(A^\dagger RAT) = 4\cos^2\left(\frac{\theta}{2}\right)[1+\cos(\alpha+\beta)] + 12, \tag{18}$$

$EP(CU) = 0$ if $\cos^2(\theta/2)[1+\cos(\alpha+\beta)] = 2$. That is, if the parameters are such that $\theta = 0$ ($2\pi$) and $\alpha + \beta = 0$ ($2\pi$) the entangling power is zero. Similarly, for the maximum value of entangling power the angles must satisfy the expression

$$\cos^2\left(\frac{\theta}{2}\right)[1+\cos(\alpha+\beta)] = 0. \tag{19}$$

The above expression gives two distinct cases namely, (i) $\theta = \pi$ for any values of $\alpha$ and $\beta$ (ii) $\theta \neq \pi$ and $\alpha + \beta = \pi$, for which $CU$ possesses the entangling power $2/9$.

### (B) Perfect Entangler

In order to calculate the local invariants, it is convenient to transform $CU$ in the Bell basis as $CU_B = Q^\dagger CU\, Q$ and hence we calculate $M(CU) = (CU_B)^T (CU_B)$. Then using Eqs. (5a) and (5b) the invariants are found to be

$$G_1 = \cos^2\left(\frac{\theta}{2}\right)\cos^2\left(\frac{\alpha+\beta}{2}\right) \tag{20a}$$

$$G_2 = 2\cos^2\left(\frac{\theta}{2}\right)\cos^2\left(\frac{\alpha+\beta}{2}\right) + 1. \tag{20b}$$

It may be noted that the local invariants of $CNOT$ are $G_1 = 0$ and $G_2 = 1$, which correspond to the geometrical point $L = [\pi/2, 0, 0]$. Since this point satisfies Eq. (7), $CNOT$ is a perfect entangler [6]. By equating the above expressions to the local invariants of $CNOT$, we have

$$\cos^2\left(\frac{\theta}{2}\right)[1+\cos(\alpha+\beta)] = 0. \tag{21}$$

That is, $CU$ gates satisfying the above condition are locally equivalent to $CNOT$, and they correspond to the same point $L$. In other words, the $CU$ satisfying Eq. (21) are locally equivalent class of $CNOT$ and hence they are also perfect entanglers, as shown in appendix B.

It is interesting to note that Eqs. (19) and (21) are identical, implying that $CU$ gates which are locally equivalent to $CNOT$ must necessarily possess the maximum entangling power.

## V. DISCUSSION

In this paper we have studied the entangling character of two qubit gates namely, $SWAP^{1/m}$ and controlled unitary, using entangling power and perfect entanglers as tools. The first part of the investigation shows that $SWAP^{1/m}$ family lies along one edge ($OA_3$) of the geometrical representation of non-local two qubit gates. It is also observed that, $SWAP^{1/2}$ possesses the maximal entangling power as well as the *only* perfect entangler in the family. Further, from the entangling power point of view, it is conjectured that $CNOT$ can also be constructed using two $SWAP^{1/3}$ gates. The possibility of such a construction is left for future investigation.

In later part of the paper, we have addressed the entangling properties of controlled unitary *(CU)* which is an important class of two qubit gates. In particular, without loss generality, a simplified expression for the entangling power of two qubit gate is presented. This simplification facilitates to obtain condition on $CU$ to possess the minimum and maximum entangling power. Further, a subset of $CU$ which is locally equivalent to $CNOT$ is explicitly identified. We refer the subset as a $CNOT$ class. Interestingly, the $CNOT$ class is shown to possess the entangling power of $CNOT$, which is the maximum value. This result provokes to conjecture that, locally equivalent gates will have the same entangling power, which warrants a detailed study. We also note that, since $CNOT$ is a perfect entangler, all its locally equivalent gates are also perfect entanglers.

It is well known that an arbitrary $CU$ gate can be constructed using two $CNOT$ and single qubit gates [15]. Since the subset $CU$ and $CNOT$ are locally equivalent, in principle it is possible to construct an element of the subset with single $CNOT$. Such a construction would be of fundamental importance in the circuit complexity.

**Appendix A**

Here we present a simple technique to explicitly show that $SWAP^{1/2}$ is the only perfect entangler. Consider two single qubit states as $|\psi_1\rangle = a|0\rangle + b|1\rangle$ and $|\psi_2\rangle = e|0\rangle + f|1\rangle$ and denoting $|\eta\rangle = SWAP^{1/m}|\psi_1\rangle \otimes |\psi_2\rangle$, the concurrence can be calculated using Eq. (3) as

$$C(\eta) = 2\left|-\frac{1}{4}\left(1 - \exp\left(\frac{2\pi i}{m}\right)\right)(af - be)^2\right|$$

From the above expression, we observe that $C(\eta) = 1$ only for $m = 2$ with appropriate choices of input states like (A) $a = 0, b = 1, e = 1, f = 0$, (B) $a = e = \frac{1}{\sqrt{2}}, b = \mp \frac{1}{\sqrt{2}}, f = \pm \frac{1}{\sqrt{2}}$.

**Appendix B**

As seen from Eq. (19) $CU$ is a perfect entangler for (i) $\theta = \pi$ for any values of $\alpha$ and $\beta$ (ii) $\theta \neq \pi$ and $\alpha + \beta = \pi$. Following the earlier technique, here we show that the $CU$ can generate maximally entangled state for some input states.

(i) $\theta = \pi$ for any values of $\alpha$ and $\beta$:

Adopting the notations used in appendix A, we denote $|\eta\rangle = CU |\psi_1\rangle \otimes |\psi_2\rangle$ and the concurrence is

$$C(\eta) = 2\left|ab\left\{e^2 \exp\left(-i\left(\frac{\alpha - \beta}{2}\right)\right) - f^2 \exp\left(i\left(\frac{\alpha - \beta}{2}\right)\right)\right\}\right|.$$

It is easy to verify that $C(\eta) = 1$ for the following choices of input states:

(A) $a = \frac{1}{\sqrt{2}}, b = \pm \frac{1}{\sqrt{2}}, e = 1, f = 0$ (B) $a = \frac{1}{\sqrt{2}}, b = \pm \frac{1}{\sqrt{2}}, e = 0, f = 1$.

(ii) $\theta \neq \pi, \alpha + \beta = \pi$.

In this case the concurrence can be obtained as

$$C(\eta) = 2\left|-2i\, abef \cos\left(\frac{\theta}{2}\right) - i\, ab \sin\left(\frac{\theta}{2}\right)\left(e^2 \exp(-i\alpha) - f^2 \exp(i\alpha)\right)\right|.$$

As an example, we can easily verify that for an arbitrary value of $\alpha$ and $\theta=0$, the application of $CU$ on the input state $a=b=e=\frac{1}{\sqrt{2}}, f=\pm\frac{1}{\sqrt{2}}$ produces maximally entangled state.